\documentclass[aps,prb,twocolumn]{revtex4}

\usepackage{graphicx}
\begin{document}

\author{Benoy Chakraverty \\
Lepes, C.N.R.S, Grenoble 38042, France.}
\title{A phenomenological model of the Resonance peak in High T$_{c}$
Superconductors}

\begin{abstract}
A notable aspect of high-temperature superconductivity in the copper oxides
is the unconventional nature of the underlying paired-electron states. The
appearance of a resonance peak, observed in inelastic neutron spectroscopy
in the superconducting state of the High T$_{c}$\ cuprates, its apparent
linear correlation with the critical superconducting temperature of each of
the compounds and its disappearance in the normal state are rather
intriguing. It may well be that this peak is the signature of the singlet to
triplet excitation, and is an unique characteristic of a d-wave
superconductor. We develop a simple criterion for the resonance peak which
is based on the concept of twist stiffness and its disappearance at T=T$_{c}$.
\end{abstract}
\date{20/06/2005}
\maketitle
The most notable feature of the unconventional nature of the High $T_{c}$
cuprates besides its near-neighbor singlet ground state, is its
superconducting gap $\Delta $; unlike conventional $B.C.S$ behavior, where $%
\Delta$, the ampltude of the gap goes up as $T_{c}$ goes up, the measured $%
\Delta $ as revealed by angular resolved photoemission spectroscopy\cite
{ding} $goes$ $down$ as $T_{c}$ goes up! This had led some authors \cite
{emery}to postulate that the energy scale governing $T_{c}$ is phase
stiffness of the order parameter or the superfluid density rather than the
modulus of $\Delta $ where $\Delta ^{*}\Delta $ is the superconducting
condensate density. Inelastic neutron scattering in High T$_{c}$ cuprate
compounds has been of immense help to enhance our understanding of the
magnetic aspects $underlying$ physics of High T$_{c}$.\cite{regnault} It
told us right away without any ambiguity that there are at least two clear
signatures of the unconventional superconductivity: spin gap and resonance
frequency. Some general features emerge from all the compunds so far studied:

(a) Local antiferromagnetic or singlet correlations in the normal and
superconducting states are observed , as evidenced by an incoherent
background of spin excitation, $S\left( q,\omega \right)$, particularly at
wave vector $q=\pi ,\pi $ \& frequency $\omega$.

(b) On the low energy side , an excitation energy gap , called spin gap $
E_{g}$ opens up in the superconducting state, which tends to zero at the
critical hole concentration where superconductivity first appears and is a
maximum at optimum doping \cite{rossatmignod}.

(c) The most unexpected feature of the inelastic neutron spectroscopy is the
emergence of an extremely intense and narrow peak $only$ in the
superconducting state at the resonance energy $\omega _{r},$ at $q=\pi ,\pi $
that is a hallmark of each superconducting compound. ($YBa_{2}Cu_{3}O_{7-x}$,
\cite{siddis} $Bi_{2}Sr_{2}Ca Cu_{2}O_{8+\delta }$\cite{fong},$
La_{2-x}Sr_{x}CuO_{4}$\cite{lake}). This resonance is a collective spin
excitation mode where the magnetic excitation spectrum condenses into a peak
at a well defined energy. It generally disappears when $T_{c}$ goes to zero
and is a generic feature of all the cuprates. The striking characteristic of
the resonance peak is its $linear$ scaling with $T_{c},$ as measured for a
variety of dopings \cite{bourges}.

Our main objective in this communication is to convey an underlying
universality relating to the resonance peak; the simplicity of the model and
its connection to underlying symmetries is its appealing feature.

We assume to start with that the superconducting ground state is a d-wave
singlet. In order to bring out the underlying symmetry elements of the
superconducting and the normal state, let us introduce the well known
concept of superconducting phase stiffness (related to charge stiffness) 
\cite{kohn}, spin stiffness \cite{shastry}as well as that of twist stiffness
which is of particular relevance to near neighbor singlets and is associated
with chirality. Each of these three stiffnesses are associated with a
distinct symmetry operation and expresses the energy increase of the system
as each symmetry operation is applied. Let us consider a spinor on site $i$ 
\begin{equation}
\psi _{i}=\left( 
\begin{array}{c}
c_{i\uparrow }^{\dagger } \\ 
c_{i\downarrow }^{\dagger }
\end{array}
\right)
\end{equation}
Here the $c_{i\uparrow }^{\dagger }$ are the electron creation operator on
site $i$ in a spin state $\uparrow $ and similarly for the other spin $%
\downarrow $. There are three sets of transformation that we can consider on
the spinors, one in the charge sector,one in the spin sector and one in the
twist sector.

(a) In the charge sector it is given by 
\begin{equation}
\psi _{i}^{^{\prime }}=\exp \left( ie\varphi _{i}\right) \psi _{i}
\end{equation}
where $e$ $is$ the electron's charge causing a rotation by an angle $\varphi
_{i}$, in the electromagnetic gauge space. This is the one parameter
transformation of symmetry group $U(1)$. In any superconducting ground
state, the $U(1)$ symmetry will be broken signifying blocking of the phase $%
\varphi $ of the superconducting order parameter and hence a non zero
superconducting phase stiffness $D_{s}.$ \cite{scalapino}

(b) We can also rotate the spinor in the spin sector by rotating the spin
through an angle $\theta _{i}$ around the spin $\sigma -axis$ so that 
\begin{equation}
\psi _{i}^{^{\prime }}=\exp \left( i\frac{\sigma }{2}\theta _{i}\right) \psi
_{i}
\end{equation}
where $\sigma $ $is$ the Pauli spin matrix. The group symmetry is $SO(3)$ or 
$SU(2)$. We note that if the ground state is a superconducting $d-$spin
singlet $S=0,$ the ground state energy will be unaffected by rotation of the
spin axis ,whatever the Hamiltonian is and as a result the spin stiffness $%
D_{\sigma }$, is necessarily zero $.$

(c) The twist stiffness is best understood by introducing the chirality
where we write 
\begin{equation}
\psi _{i}^{^{\prime }}=\exp \left( \frac{i\sigma \text{ }\gamma _{5}\theta
_{i}}{2}\right) \psi _{i}
\end{equation}
here the chirality operator $\gamma _{5}$\cite{sakurai}transcribes the fact
that the spin rotation $\theta _{i}$ on site $i$ is exactly equal and
opposite to that on the near neighbor site $j$ whence $\theta _{i}-\theta
_{j}=2\theta .$ This gives 
\begin{eqnarray}
\psi _{i}^{^{\prime }} &=&\exp \left( \frac{i\sigma \text{ }\theta }{2}%
\right) \psi _{i}\text{ ;} \\
\psi _{j}^{^{\prime }} &=&\exp \left( \frac{-i\sigma \text{ }\theta }{2}%
\right) \psi _{j}  \nonumber
\end{eqnarray}
If the site $i$ and $j$ belong to sublattice $A$ and $B,$ then the chiral
rotation twists one sublattice around another by a rigid angle $\theta .$
The symmetry of the operation because of two sites is $SU(2)\times SU(2)$
which is in the same homotopy class as $SO(4).${\it If the ground state is a
near neighbor singlet ,the twist rotation }$\theta ${\it \ mixes the singlet
with the triplet and hence leads to increase of the ground state energy}.
This will be clear if we consider the four basis pair states on near
neighbor sites $i$ \& $j$ written as \cite{sachdev} 
\begin{eqnarray}
&\mid &b\rangle =\frac{1}{\sqrt{2}}\left( c_{i\uparrow
}^{+^{{}}}c_{j\downarrow }^{+^{{}}}-c_{i\downarrow }^{+^{{}}}c_{j\uparrow
}^{+^{{}}}\right) \mid 0\rangle   \nonumber \\
&\mid &t_{x}\rangle =\frac{-1}{\sqrt{2}}\left( c_{i\uparrow
}^{+}c_{j\uparrow }^{+}-c_{i\downarrow }^{+}c_{j\downarrow }^{+}\right) \mid
0\rangle   \nonumber \\
&\mid &t_{y}\rangle =\frac{i}{\sqrt{2}}\left( c_{i\uparrow }^{+}c_{j\uparrow
}^{+}+c_{i\downarrow }^{+}c_{j\downarrow }^{+}\right) \mid 0\rangle  
\nonumber \\
&\mid &t_{z}\rangle =\frac{1}{\sqrt{2}}\left( c_{i\uparrow
}^{+}c_{j\downarrow }^{+}+c_{i\downarrow }^{+}c_{j\downarrow }^{+}\right)
\mid 0\rangle 
\end{eqnarray}
Here $\mid b\rangle $ $=b^{\dagger }\mid 0\rangle $ \& $\mid t_{\alpha
}\rangle =t_{\alpha }^{\dagger }\mid 0\rangle .$ $\mid b\rangle $ is a $S=0$
singlet while the three $\mid t_{\alpha }\rangle $ are $S=1,$ triplets and
the four states constitute the symmetry $SO(4)$. Effect of chiral rotaion on
site $i\&$ $j$ with these basis wave functions. will give $4\times 4$ matrix 
\begin{equation}
\left( 
\begin{array}{c}
b \\ 
t_{\alpha }
\end{array}
\right) ^{^{\prime }}=\exp \left( i\sigma _{\alpha }\theta \right) \left( 
\begin{array}{c}
b \\ 
t_{\alpha }
\end{array}
\right) 
\end{equation}
The twist $\theta $ has mixed up singlets and triplets. This makes the twist
stiffness $D_{t}$ $\neq 0$ (the twist operation is nonunitary in this ground
state). The Lie algebra of $SO(4)$ is closed by the three generators that
connect $t_{\alpha }$ amongst themselves by pure spin axis
rotation,expression (3) and by the other three generators that connect $b$
with the three $t$ 's through the twist rotation (5). As for illustration
let us a take a model hamiltonian $t-J$ hamiltonian\cite{zhang} 
\begin{equation}
H_{o}=-t\sum_{i,j}c_{i\sigma }^{\dagger }c_{j\sigma }-J\sum_{i,j}^{i\neq %
j}S_{i}S_{j}
\end{equation}
Here the first term is the electron hopping between sites $i$ and $j,$ $t$
being the hopping integral while the last term is a Heisenberg
antiferromagnetic exchange interaction $J$ between first near neighbor spins 
$S_{i}$ and $S_{j}$ on site $i$ and $j.$Does $t-J$ hamiltonian have a
superconducting ground state? The first of the gauge transformation
(expression $2$ ) has been used to show that its sibling, the Hubbard
hamiltonian (in the large $U$ limit) has a superconducting ground state with
a nonzero phase stiffness $D_{s}$ at $T=0$\cite{mohit}$.$ What about the
twist stiffness of the Hamiltonian $(8)$? In order to calculate the twist
stiffness, we apply uniform twist $\theta ,$ between near neighbor sites $i$
\& $j$ of sublattice $A$ and $B$. It is convenient to transform $H_{o}$ in
terms of singlet and triplet pair operators using the pair representation of
the spin operators\cite{sachdev}which we write as 
\begin{eqnarray}
\nonumber H_{o}&=&-t\sum_{i,j}c_{i\sigma }^{\dagger }c_{j\sigma }-\frac{3J}{4}\sum_{\nu
}b_{\nu }^{\dagger }b_{\nu }+\frac{J}{4}\sum_{\nu ,\alpha }t_{\nu \alpha
}^{\dagger }t_{\nu \alpha }\\
&&-\mu \sum_{i}\left[ b_{\nu }^{2}+t_{\nu
}^{2}\right] 
\end{eqnarray}

Here $\mu $ is the chemical potential assumed same over all space. This term
is unaffected by twist and we assume that the sum $\sum_{\nu }\left[ b_{\nu
}^{2}+t_{\nu }^{2}\right] $ over $\nu $ near neighbor pairs which is $%
=N_{electron}$ is conserved. In the limit of small twist the Hamiltonian
gets modified 
\begin{equation}
H^{^{\prime }}=H\left( 0\right) +H\left( \theta \right) 
\end{equation}
where the first part is the unperturbed untwisted Hamiltonian. By developing 
$H\left( \theta \right) $ to second order we obtain for the perturbing term 
\begin{equation}
H\left( \theta \right) =\sum_{i,j}\left[ j_{ij}^{\sigma }\theta -\frac{1}{4}%
T_{ij}\theta ^{2}\right] 
\end{equation}
where $j_{ij}^{\sigma }$ is the spin current operator and $T_{ij}$ is the
kinetic energy operator.\ They are given respectively by 
\begin{eqnarray}
\hat{j}&=&it\left[\sum_{ij}\left( c_{i\sigma }^{\dagger }\sigma \gamma
_{5}c_{j\sigma }-H.C.\right) -iJ \sum_\nu\left( b_{\nu }^{\dagger }t_{\nu \alpha
}-t_{\nu \alpha }^{\dagger }b_{\nu }\right)\right] \nonumber \\
\hat{T}&=&-t\sum_{ij}\left( c_{i\sigma }^{\dagger }c_{j\sigma }+H.C\right) -\frac{J}{2}%
\sum_\nu\left( b_{\nu }^{2}+t_{\nu }^{2}\right) 
\end{eqnarray}
We get ground state energy shift due to twist $\theta $ as 
\begin{equation}
\Delta E_{o}=\left\langle H\left( \theta \right) \right\rangle =\frac{N}{2}%
D_{t}\theta ^{2}
\end{equation}
$D_{t}\left( \omega =0\right)$ is the twist stiffness (in two dimensions it
has the dimension of energy).\ It is formally given by 
\begin{equation}
D_{t}\left( \omega \right) =\frac{1}{N}\left[ \left\langle
-\hat{T}\right\rangle -\sum_{n\neq 0}\left( \frac{\left\langle 0\mid
\hat{j}^{\sigma }\mid n\right\rangle ^{2}}{\epsilon _{n}-\epsilon _{o}-\hbar
\omega }-\frac{\left\langle n\mid \hat{j}^{\sigma }\mid 0\right\rangle ^{2}}{%
\epsilon _{o}-\epsilon _{n}-\hbar \omega }\right) \right] 
\end{equation}
In the absence of the hopping term and of the spin current term, the energy
increase per electron is precisely $J$ which is the bare twist stiffness.
The first term of $D_{t}\left( \omega \right) $ is the diamagnetic current
contribution to stiffness due to the average value of the kinetic energy
while the second term reflects second order contribution of $``$
paramagnetic spin current conductivity$^{"}\sigma _{t}(\omega )$ although $%
\left\langle j^{\sigma }\right\rangle =0$. The energy levels $\epsilon _{n}$
are the triplet excited states for a momentum transfer $\pi ,\pi $ (which
has a gap $E_{g}$ as measured by inelastic neutron spectroscopy). The spin
current in the twisted frame is the response to a ``twist vector potential''
(engendered by local twist) $just$ as the charge current is response to an
electromagnetic vector potential. The linear coefficient of the total
response is the corresponding twist stiffness. We can rewrite the expression
(14) more conveniently in analogy to the missing area sum rule\cite{tinkham}%
of the missing Drude weight as 
\begin{equation}
D_{t}\delta (\omega )=D_{t}^{o}\delta (\omega )-\int_{o}^{\infty }\sigma
_{t}(\omega )d\omega 
\end{equation}
Here the second term on the right reflects the exhaustion of twist rigidity
through incoherent spin excitation where $\sigma _{t}(\omega )$ $is$ $\sim 
Im\chi _{\bot }\left( \omega \right) ,$ the trasverse spin
susceptibility. From the experimental neutron data \cite{bourges}, we know
that ${Im}\chi _{\bot }\left( \omega \right) is$ very large at the
critical hole concentration $\partial _{h}^{c}$ at which $T_{c}=0$ while $%
{Im}\chi _{\bot }\left( \omega \right) $ monotonically decreases (
integrated spectral weight) in the superconducting state as optimum doping $%
\partial _{h}^{opt}$is approached so that we can reasonably conclude that $%
D_{t}=0$ at $\partial _{h}=\partial _{h}^{c}$ while $D_{t}$ ought to be a
maximum at $\partial _{h}=\partial _{h}^{opt}.$In otherwords $D_{t}$ is a
correct indicator of d-wave superconductivity. The non-zero phase stiffness
in conventional $s-wave$ superconductor results from broken $U(1)$
electromagnetic gauge symmetry. The non-zero spin stiffness in a system with
long range magnetic order is associated with a broken $SO(3)$ symmetry of
the rotational invariance of the spin space and $D_{\sigma }$ goes to zero
at $T=T_{N}$ when the invariance is restored. What symmetry or symmetries
are broken when the phase coherent singlet d-wave ground state emerges? We
may think of the $d-wave$ superconducting state as a state where $SO(4)$
symmetry is explicitly broken as well as $U(1)$. The normal state is then a
state with zero twist stiffness where the broken $SO(4)$ symmetry pertaining
to singlet and the three triplets has been restored. If now we accept the
premise that at $T_{c}$, twist stiffness $D_{t}$ goes to zero,then one makes
the simple statement that $kT_{c}$ is equal to the value of twist stiffness
at $T=0$ (strictly speaking one should use renormalised stiffness due to
triplet excitations) and we have 
\begin{equation}
kT_{c}=D_{t}\left( T=0,\omega =0\right) 
\end{equation}
The expression relating spin stiffness to some characteristic frequency
(which we shall baptise resonance frequency $\omega _{r}$) can be written as 
\cite{halperin}\cite{legett}\ 
\begin{equation}
D_{t}\left( T,\omega =0\right) =\chi _{\perp }\left( T\right) \hbar
^{2}\omega _{r}^{2}\left( T\right) 
\end{equation}
The resonance frequency $\omega _{r}$ is a small amplitude harmonic twist
oscillation or rigid precession of sublattice $A$ with respect to sublattice 
$B.$ Here $\chi _{\perp }\left( T\right) $ is the transverse spin flip
magnetic susceptibility, which has its largest value at $Q=\pi ,\pi $. The
transverse static susceptibility $\chi _{\perp }\left( T\right) $ in the
High $T_{c}$ cuprates (as measured by N.M.R $\frac{1}{T_{2G}}$ spin -spin
relaxation rate) can be parametrised as \cite{julien} 
\begin{equation}
\chi _{\perp }\left( T\right) =\frac{A}{k\left( T+T_{c}\right) }
\end{equation}
where $A$ is a phenomenogical constant. That this form of the static
susceptibility in the normal state at $T\gg T_{c}$ is appropriate can be
checked from the imaginary part of susceptibility 
\begin{equation}
{Im}\chi _{\perp }\left( \omega ,T\right) \sim \frac{\omega }{T}
\end{equation}
which is of a form universally observed for small $\frac{\omega }{T}$\cite
{keimer}. This behavior in the normal state probably points to proximity to
a quantum critical point for spin excitation. In a temperature range above $%
T_{c}$ the spin correlations have a rapid decay in space but a slow decay in
time due to a large density of $S=1$ excited states. Real part of the
dynamical susceptibility $\chi _{\perp }\left( q,\omega \right) $ would not
show a narrow peak around a specific ordering vector but ${Im}\chi
_{\perp }\left( T,\omega \right) $ will exhibit considerable weight at low
frequency. Using expression (17) and (18) ,we obtain 
\begin{eqnarray}
&&  \nonumber \\
\hbar \omega _{r} &=&akT_{c}
\end{eqnarray}
\begin{figure}[htbp]
  \centering
  \includegraphics[scale=0.6,angle=0]{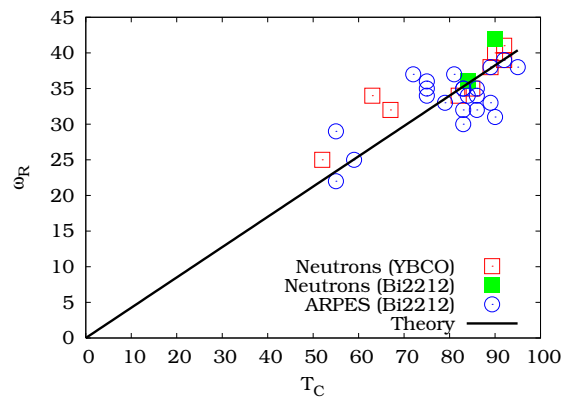}
  \caption{Resonance frequency vs. superconducting critical temperature $T_c$,
 extracted from references [5,6,8,21]. }
\end{figure}
This is our central result. It corroborates a posteriori the central
assumpion that the coherent part of the spectral weight is very simply
related to twist stiffness and emerges as coherent resonance peak; and whose
intensity reflects the incoherent spin excitations that have disappeared
from the energy range $0\leq \omega \leq E_{g}$ in accordance with the sum
rule (15). Superconductivity can only arise as the $d-wave$ singlets manage
to shake off the triplets from the normal state soup of singlets\& triplets,
as a result of the opening up of the spin gap $E_{g}\left( \pi ,\pi \right) $%
. The expression (23) is plotted in figure (1), with neutron and ARPES\cite
{campuzano} data superimposed. The proportionality constant $a$ measured
from fig 1 gives the number $0.42$ $me\upsilon /{{}^{\circ }}K$. If we are
at the critical hole doping concentration $\partial _{h}^{c}$ at which both $%
T_{N}$,the Neel temperature $\&$ $T_{c},$ the superconducting critical
temperature are both zero, then we must have $D_{\sigma }=0$ and $D_{t}=0,$
signifying no long range magnetic order and no long range superfluid order;
it is a quantum critical point. It is well known that $Zn$ doping destroys $%
T_{c}.$ It is seen by neutrons that doping with $Zn$ introduces large low
energy spin fluctuations(integrated spectral weight increaes,the spin gap $%
E_{g}(\pi ,\pi )$ rapidly goes to zero), that will drive $D_{t}$ to zero
suppressing $\omega _{r}$ and killing superconductivity. The normal state
can be defined as a spin liquid (by definition has no sublattice
magnetisation) where we have considerable low energy spin excitation. We
also require that translational invariance be unbroken for the system to
qualify as a liquid. Thus it describes a gapless spin liquid more in
conformity with the original suggestion of the long range RVB liquid \cite
{anderson} In its loss of twist stiffness the spin liquid behaves like any
conventional liquid loosing shear rigidity at the melting transition. The
concept of twist stiffness is based on $infinitesimally$ $small$ $twist$ as
is customary in these definitions; beyond $T\succeq T_{c},$ the restored
dynamical $SO(4)$ symmetry implies $b\Longleftrightarrow t_{\alpha }$ pair
fluctuation in the spin liquid phase costing no energy around the untwisted
singlet. If this symmetry persists for $all$ twist angles then we will be in
the frustrated ``Henley limit''\cite{henley} of infinite classical spin
degeneracy where one sublattice $A$ will twist freely around the other
sublattice $B$ and the two sublattices are totally decoupled. Or else the
system may develop a region where $D_{t}\left( T\succeq T_{c}\right) $ $may$ 
$become$ $negative$ for $large$ $twist$ $angles$ generating large
singlet-triplet excursions and hence may go spontaneously to a distorted or
twisted ground state \cite{siggia}. Although twist stiffness and
superconducting phase stiffness are different at $T=0,$ their simultaneous
disappearance at $T=T_{c}$ is indicated by the Arpes results \cite{campuzano}%
of the hump and dip structure in the electronic spectral weight and point to
strong coupling of triplet and phase fluctuation as $T_{c}$ is approached.

Several theoretical models exist \cite{maki} that explain the resonance
peak. Our objective in this paper has been relatively simple: can we
understand the resonance peak without a detailed model and does it have some
predictive ability as to the underlying symmetry nature of the normal and
superconducting state? I think the arguments given in this paper will throw
some new light on these issues.

The author acknowledges highly stimulating discussions with Dr M. Cuoco,
He benefited from comments by  M. Avignon,  
P. Bourges, T. Chatterjee, K. Jain, J. Ranninger, T. Ziman, 
The author is very grateful to Dr S. Fratini 
for a careful reading of the manuscript at various stages.


\begin{thebibliography}{99}
\bibitem{ding}  H.Ding et al., Nature (london) {\bf 382, 51 (1996)};
J.M.Harris et al, Phys Rev {\bf B 54 }, R15665 (1996).

\bibitem{emery}  V.J. Emery \& S.A. Kivelson, Nature {\bf 374, }434 (1995);
B.K. Chakraverty, A. Taraphder \& M. Avignon, Physica C {\bf \ 235-240 },
2323 (1994), B.K.Chakraverty\& T.V.Ramakrishnan, Physica {\bf \ C282},
290(1997).

\bibitem{regnault}  L.P.Regnault, Ph. Bourges \& P.Burlet -` {\it Neutron
Scattering in LayeredCopper Oxydes Superconductors' Ed: A. Furer (Kluiver
Academic Publications, 1998)} p. 65.

\bibitem{rossatmignod}  J.Rossad-Mignod et al., Physica B {\bf 169 }, 58
(1991); Physica C {\bf 185-189 }, 86 (1991).

\bibitem{siddis}  P.Bourges et al., Phys Rev B {\bf 53 }, 876 (1996); P. Dai
et al, Phys. Rev. Lett. {\bf 77}, 5425 (1996); H.F. Fong et al, Phys. Rev. Lett. 
{\bf 78}, 713 ( 1997 ).

\bibitem{fong}  H.F. Fong et al, Nature (London) {\bf 398}, 588 (1999).

\bibitem{lake}  B. Lake et al, Nature (London) {\bf 400}, 43 (1999).

\bibitem{bourges}  P. Bourges, ` {\it The gap Symmetry \& Fluctuations
in High T}$_{c}${\it \ Superconductors }' (Ed: J.Bok et al., Plenum Press,
N.Y, 1998) p. 85; cond-mat/{ 9901333}.

\bibitem{kohn}  W. Kohn, Phys Rev{\bf \ 133}, A171 (1964)

\bibitem{shastry}  B. Sriram Shastry \& B. Sutherland - Phys. Rev. Lett. {\bf %
65 }, 243 (1990).

\bibitem{scalapino}  D.J. Scalapino, S.R. White \& S.C. Zhang - Phys. Rev.
Lett.  {\bf 68}, 2830 (1992)

\bibitem{sakurai}  J.J. Sakurai, {\it ''Advanced quantum mechanics''}
(Benjamin Publishing, California, U.S.A 1984) p. 169.

\bibitem{sachdev}  S. Sachdev \& R.N. Bhatt, Phys. Rev. {\bf B 41}, 9323 (1990).

\bibitem{zhang}  F.C. Zhang\& T.M. Rice, Phys. Rev. {\bf B37}, 3759 (1988).




\bibitem{mohit}  A. Paramekanti, M. Randeria and N. Trivedi, Phys.
Rev. Lett.  {\bf 87}, 217002 (2001)

\bibitem{tinkham}  R.E. Glover \& M. Tinkham, Phys. Rev. {\bf 104}, 844 (1956)



\bibitem{halperin}  B.I. Halperin \& P.C. Hohenberg - Phys. Rev. {\bf 177},
952 (1969).

\bibitem{legett}  A. J. Legett - Nobel Prize Lecture, 2003, Rev Mod
Phys, {\bf 76}, 999 (2004).

\bibitem{julien}   Y. Itoh et al, J. Phys. Soc. Japan {\bf %
61 }1287 (1992); M.-H. Julien, - {\it Thesis, University of
Grenoble, France} 1997, p 51
\bibitem{keimer}  B. Keimer et al, Phys. Rev. Lett. {\bf 67} 1930 ( 1991)

\bibitem{campuzano}  J.C.Campuzano et al, Phys. Rev. Lett. {\bf 83}, 3709
(1999).

\bibitem{anderson}  P.W. Anderson, G. Baskaran, Z.Zou \&T.Hsu, Phys. Rev. Lett. 
{\bf 58 }, 2790 (1987).

\bibitem{henley}  C. L. Henley, Phys. Rev. Lett. {\bf 62}, 2056 (1989).

\bibitem{siggia}  B. Shraiman \& E. D.Siggia, Phys. Rev. Lett. {\bf 62 },
1564 (1989).

\bibitem{maki}  K. Maki \& H. Won, Phys. Rev. Lett. {\bf 72}, 1758 (1993); 
E.\ Demler \& S.C. Zhang, Phys. Rev. Lett. {\bf 75}, 4126 (1995); Chien-hua
Pao \& N.E. Bickers, Phys Rev B {\bf \ 51 }, 16310 (1995); N. Bulut \& D.J.
Scalapino, Phys. Rev. {\bf 53}, 5149 (1996); E. Altman \& A. Auerbach,
Phys. Rev. B {\bf 65}, 104508 (2002).
\end{thebibliography}
\end{document}